\documentclass[%
 reprint,
superscriptaddress,
 amsmath,amssymb,
 aps,
]{revtex4-1}

\usepackage{graphicx}
\usepackage{dcolumn}
\usepackage{bm}

\begin{document}

\preprint{APS/123-QED}

\title{Long-distance continuous-variable quantum key distribution \\ using non-Gaussian state-discrimination detection}

\author{Qin Liao}%
\affiliation{School of Information Science $\&$ Engineering, Central South University, Changsha 410083, China}
\author{Ying Guo}
 \email{Corresponding author: yingguo@csu.edu.cn}
\affiliation{School of Information Science $\&$ Engineering, Central South University, Changsha 410083, China}
\author{Duan Huang}
\affiliation{School of Information Science $\&$ Engineering, Central South University, Changsha 410083, China}
\author{Peng Huang}
\affiliation{State Key Laboratory of Advanced Optical Communication Systems and Networks, and Center of Quantum Information Sensing and Processing, Shanghai Jiao Tong University, Shanghai 200240, China}
\author{Guihua Zeng}
\affiliation{State Key Laboratory of Advanced Optical Communication Systems and Networks, and Center of Quantum Information Sensing and Processing, Shanghai Jiao Tong University, Shanghai 200240, China}
\date{\today}

\begin{abstract}
We propose a long-distance continuous-variable quantum key distribution (CVQKD) with four-state protocol using non-Gaussian state-discrimination detection. A photon subtraction operation, which is deployed at the transmitter, is used for splitting the signal required for generating the non-Gaussian operation to lengthen the maximum transmission distance of CVQKD. Whereby an improved state-discrimination detector, which can be deemed as an optimized quantum measurement that allows the discrimination of nonorthogonal coherent states beating the standard quantum limit, is applied at the receiver to codetermine the measurement result with conventional coherent detector. By tactfully exploiting multiplexing technique, the resulting signals can be simultaneously transmitted through an untrusted quantum channel, and subsequently sent to the state-discrimination detector and coherent detector respectively. Security analysis shows that the proposed scheme can lengthen the maximum transmission distance up to hundreds of kilometers. Furthermore, by taking finite-size effect and composable security into account we obtain the tightest bound of the secure distance, which is more practical than that obtained in the asymptotic limit.
\end{abstract}

\pacs{Valid PACS appear here}
\maketitle


\section{\label{sec:level1} introduction}

Quantum key distribution (QKD) \cite{Bennett1984Quantum,r31,r32} is one of the most practical applications of quantum cryptography, whose goal is to provide an elegant way that allows two remote legitimate partners, Alice and Bob, to establish a sequence of random secure key over insecure quantum and classical channels. Its security is provided by the laws of quantum physics \cite{Wootters1982A,PhysRevD.74.125012}.

For decades, continuous-variable (CV) QKD \cite{Pirandola2015High,PhysRevA.89.042335,PhysRevLett.95.180503,Chen:2016dl,PhysRevA.89.042330,Zhang:2012jc,Leverrier:2011cu,Leverrier:2009cr} has been becoming a hotspot of QKD research due to its simple implementation with state-of-art techniques \cite{PhysRevLett.93.170504,PhysRevLett.88.057902}. It has been shown to be secure against arbitrary collective attacks, which are optimal in both the asymptotic limit \cite{PhysRevLett.94.020504,PhysRevLett.94.020505,Leverrier:2009cr,Leverrier:2011cu} and the finite-size regime \cite{PhysRevLett.109.100502,PhysRevLett.110.030502}. Recently, CVQKD is further proved to be secure against collective attacks in composable security framework \cite{Leverrier:2015he}, which is the security analysis by carefully considering every detailed step in CVQKD system.

In general, there are two main modulation approaches in CVQKD, i.e.,\ Gaussian modulated CVQKD \cite{PhysRevLett.95.180503,PhysRevA.89.042335,Pirandola2015High} and discretely modulated CVQKD \cite{Chen:2016dl,PhysRevA.89.042330,Zhang:2012jc, Leverrier:2011cu, Leverrier:2009cr}. In the first approach, the transmitter Alice usually continuously encodes key bits in the quadratures ($\hat{x}$ and $\hat{p}$) of optical field with Gaussian modulation \cite{PhysRevLett.97.190503}, while the receiver Bob can restore the secret key through high-speed and high-efficiency coherent detector (i.e., homodyne or heterodyne detector) \cite{PhysRevLett.88.057902,PhysRevA.89.052301}. This scheme usually has a repetition rate higher than that of single-photon detections so that Gaussian modulated CVQKD could potentially achieve higher secret key rate, whereas it seems unfortunately limited to much shorter distance than its discrete-variable (DV) counterpart \cite{Leverrier:2011cu}. The key problem is that the reconciliation efficiency $\beta$ is quite low for Gaussian modulation, especially in the long-distance transmission. To solve this problem, one has to design a perfect error correcting code which is more suitable than LDPC code at very low signal-to-noise ratio (SNR). However, this kind of error correcting code is relatively hard to design and implement. Fortunately, there exist another way to well solve the problem, that is, using discrete modulation such as the four-state CVQKD protocol, proposed by Leverrier \emph{et al}. \cite{Leverrier:2009cr}. This discretely modulated CVQKD generates four nonorthogonal coherent states and 
exploits the sign of the measured quadrature of each state to encode information rather than uses the quadrature $\hat{x}$ or $\hat{p}$ itself. This is the reason that the sign of the measured quadrature is already the discrete value to which the most excellent error-correcting codes are suitable even at very low SNR. Consequently, the four-state CVQKD protocol has the merits of both high reconciliation efficiency in the long-distance transmission and the security proof of CVQKD so that it could improve the maximal transmission distance of CVQKD.

Currently, photon-subtraction operation, which is a kind of non-Gaussian operation in essence, has been demonstrated theoretically and experimentally to extend the transmission distance of the CVQKD using two-mode entangled states \cite{Guo:2017ce,PhysRevA.93.012310,PhysRevA.87.012317} due to the fact that a suitable photon-subtraction operation would increase the entanglement degree of two-mode entangled state and thereby increase the correlation between the two output modes of two-mode entangled state. Since the entanglement-based (EB) scheme is equivalent to the prepare-and-measure (PM) one, this operation can  be employed practically implemented in protocols using coherent states with existing technologies.

Furthermore, although a high-speed and high-efficiency homodyne or heterodyne detector can be effectively to measure the received quantum state, the inherent quantum uncertainty (noise) still prevents the nonorthogonal coherent states from being distinguished with perfect accuracy \cite{Becerra:2013jw,Becerra:2013ix,Becerra:2011hb}. Even if the detector is ideal with perfect detection efficiency, the receiver cannot  still obtain the precise result. The conventional ideal detector can only achieve the standard quantum limit (SQL) which defines the minimum error with which nonorthogonal states can be distinguished by direct measurement of the physical property of the light, e.g.\ quadrature $\hat{x}$ or $\hat{p}$. Actually, there exists a lower error bound known as the Helstrom bound \cite{Helstrom:110988} which is allowed by quantum mechanics, and this bound can be achieved by designing excellent state-discrimination strategies. Recently, a well-behaved state-discrimination detector has been proposed to unconditionally discriminate among four nonorthogonal coherent states in QPSK modulation \cite{Becerra:2013jw}. This detector can beat the SQL by using photon counting and adaptive measurements in the form of fast feedback and thus approach or achieve the Helstrom bound. Therefore, the performance of CVQKD would be improved by taking advantage of this well-behaved state-discrimination detector.

Inspired by the afore-mentioned advantages, which have been analyzed in theory and subsequently demonstrated with simulations and experiments, in this paper, we propose a long-distance CVQKD using non-Gaussian state-discrimination detection. In stead of the traditional Gaussian modulation which continuously encodes information into both quadrature $\hat{x}$ and quadrature $\hat{p}$, the discretely-modulated four-state CVQKD protocol is adopted as the fundamental communication protocol since it can well tolerate lower SNR, leading to the long-distance transmission compared with Gaussian-modulated counterpart. Meanwhile, a photon subtraction operation is deployed at the transmitter, where it is not only used for splitting the incoming signal, but also improving the performance of CVQKD as it has been proven to be beneficial for lengthening the maximal transmission distance. Moreover, an improved state-discrimination detector is applied at the receiver to codetermine the measurement result with coherent detector. The state-discrimination detector can be deemed as the optimized quantum measurement for the received nonorthogonal coherent states so that it could surpass the standard quantum limit. As a result, one can obtain precise result of incoming signal in QPSK format with the help of the state-discrimination detector. By exploiting multiplexing technique, the yielded signals can be simultaneously transmitted  through an untrusted quantum channel, and subsequently sent to the improved state-discrimination detector and the coherent detector. The proposed long-distance CVQKD scheme can greatly increase the secure transmission distance and thus outperforms the existing CVQKD protocols in terms of the maximal transmission distance. Taking finite-size effect and composable security into account we obtain the tightest bound of the secure distance, which is more practical than that obtained in asymptotic limit.

This paper is structured as follows. In Sec. \uppercase\expandafter{\romannumeral2}, we first introduce the discretely modulated CVQKD protocols, in particular, the four-state CVQKD protocol, and then demonstrate the proposed long-distance CVQKD scheme. In Sec. \uppercase\expandafter{\romannumeral3}, we elaborate the characteristics of photon-subtraction operation and the principle of improved state-discrimination detector. Numeric simulation and performance analysis are discussed in Sec. \uppercase\expandafter{\romannumeral4}, and finally conclusions are drawn in Sec. \uppercase\expandafter{\romannumeral5}.

\section{long-distance CVQKD scheme}

We consider the four-state CVQKD protocol as a fundamental communication protocol for the proposed scheme, since the discretely-modulated protocol is more suitable for long-distance transmission (lower SNR) and it could be extended larger than its Gaussian modulation counterparts. Furthermore, the transmission distance of the four-state CVQKD protocol can be enhanced by performing a proper photon-subtraction operation and applying a well-behaved state-discrimination detector. To make the derivation self-contained, in this section, we first briefly describe the discretely modulated four-state CVQKD protocol, and then give the detail structure of the long-distance CVQKD scheme. 

\subsection{Four-state CVQKD protocol}

In general, the four-state CVQKD protocol is derived from discretely modulated CVQKD, which can be generalized to the one with $N$ coherent states $|\alpha_k^N\rangle=|\alpha e^{i2k\pi/N}\rangle$, where $k\in\{0, 1, ... , N\}$ \cite{PhysRevA.89.042330}. For the four-state CVQKD protocol, we have $|\alpha_k^4\rangle=|\alpha e^{i(2k+1)\pi/4}\rangle$, where $k\in\{0, 1, 2, 3\}$, $\alpha$ is a positive number related to the modulation variance of coherent state as $V_M=2\alpha^2$.

Let us consider the PM version of the four-state CVQKD protocol first. Alice randomly chooses one of the coherent states $|\alpha_k^4\rangle$ and sends it to the remote Bob through a lossy and noisy quantum channel, which is characterized by a transmission efficiency $\eta$ and an excess noise $\varepsilon$. When Bob receives the modulated coherent states, he can apply either homodyne or heterodyne detector with detection efficiency $\tau$ and electronics noise $v_{el}$ to measure arbitrary one of the two quadratures $\hat{x}$ or $\hat{p}$ (or both quadratures). The mixture state that Bob received can be expressed with the following form
\begin{equation}
\begin{aligned}
\rho_4=\frac{1}{4}\sum_{k=0}^3|\alpha_k^4\rangle\langle\alpha_k^4|.
\end{aligned}
\end{equation}
After measurement, Bob then reveals the absolute values of measurement results through a classical authenticated channel and keeps their signs. Alice and Bob exploit the signs to generate the raw key. After conducting post-processing procedure, they can finally establish a correlated sequence of random secure key.

The PM version of the protocol is equivalent to the EB version, which is more convenient for security analysis. In EB version, Alice prepares a pure two-mode entangled state
\begin{equation}
\begin{aligned}
|\Psi_4\rangle &=\sum_{k=0}^3\sqrt{\lambda_k}|\phi_k\rangle|\phi_k\rangle \\
&=\frac{1}{2}\sum_{k=0}^3|\psi_k\rangle|\alpha_k^4\rangle,
\end{aligned}
\end{equation}
where the states
\begin{equation}
\begin{aligned}
|\psi_k\rangle=\frac{1}{2}\sum_{m=0}^3e^{i(1+2k)m\pi/4}|\phi_m\rangle
\end{aligned}
\end{equation}
are the non-Gaussian states, and the state $|\phi_m\rangle$ is given by
\begin{equation}
\begin{aligned}
|\phi_k\rangle=\frac{e^{-\alpha^2/2}}{\sqrt{\lambda_k}}\sum_{n=0}^\infty(-1)^n\frac{\alpha^{4n+k}}{\sqrt{(4n+k)!}}|4n+k\rangle,
\end{aligned}
\end{equation}
with
\begin{equation}
\begin{aligned}
\lambda _{0,2}=\dfrac {1}{2}e^{-\alpha^2}\left[ \cosh \left( \alpha ^{2}\right) \pm \cos \left( \alpha^{2}\right) \right],
\end{aligned}
\end{equation}
\begin{equation}
\begin{aligned}
\lambda _{1,3}=\dfrac {1}{2}e^{-\alpha^2}\left[ \sinh \left( \alpha ^{2}\right) \pm \sin \left( \alpha ^{2}\right) \right].
\end{aligned}
\end{equation}
Consequently, the mixture state $\rho_4$ can be expressed by
\begin{equation}
\begin{aligned}
\rho_4&=\mathrm{Tr}(|\Psi_4\rangle\langle\Psi_4|) \\
&=\sum_{k=0}^3\lambda_k|\phi_k\rangle\langle\phi_k|.
\end{aligned}
\end{equation}
Let $A$ and $B$ respectively denote the two output modes of the bipartite two-mode entangled state $|\Psi_4\rangle$, $\hat{a}$ and $\hat{b}$ denote the annihilation operators applying to mode $A$ and $B$ respectively. We have the covariance matrix $\Gamma_{AB}$ of the bipartite state $|\Psi_4\rangle$ with the following form
\begin{equation}\label{CM4}       
\Gamma_{AB}=
\left(                 
\begin{array}{cc}     
X\mathbb{I} & Z_4\sigma_{z}\\  
    Z_4\sigma_{z} & Y\mathbb{I}\\  
  \end{array}
\right),                 
\end{equation}
where $\mathbb{I}$ and $\sigma_{z}$ represent $\mathrm{diag}(1,1)$ and $\mathrm{diag}(1,-1)$ respectively, and
\begin{equation}
\begin{aligned}
X&=\langle  \Psi _{4}\left| 1+2a^\dag a\right| \Psi _{4}\rangle  =1+2\alpha^2,\\
Y&=\langle  \Psi _{4}\left| 1+2b^\dag b\right| \Psi _{4}\rangle  =1+2\alpha^2,\\
Z_4&=\langle  \Psi _{4}\left| ab+a^{\dag}b^{\dag}\right| \Psi _{4}\rangle=2\alpha^2\sum_{k=0}^3\lambda_{k-1}^{3/2}\lambda_k^{-1/2}.
\end{aligned}
\end{equation}
Note that the addition arithmetic should be operated with modulo 4. The detailed derivation of the four-state CVQKD protocol can be found in \cite{Leverrier:2011cu}.

After preparing the two-mode entangled state $|\Psi_4\rangle$ with variance $V=1+V_M$, Alice performs projective measurements $|\psi_k\rangle\langle\psi_k| \; (k=0,1,2,3)$ on mode $A$, which projects another mode $B$ onto a coherent state $|\alpha_k^4\rangle$. Alice subsequently sends mode $B$ to Bob through the quantum channel. Bob then applies homodyne (or heterodyne) detection to measure the incoming mode $B$. Finally, the two trusted parties Alice and Bob extract a string of secret key by using error correction and privacy amplification.

\subsection{Long-distance discretely modulated CVQKD}

\begin{figure*}
\centering
\includegraphics[width=6.2in,height=2.15in]{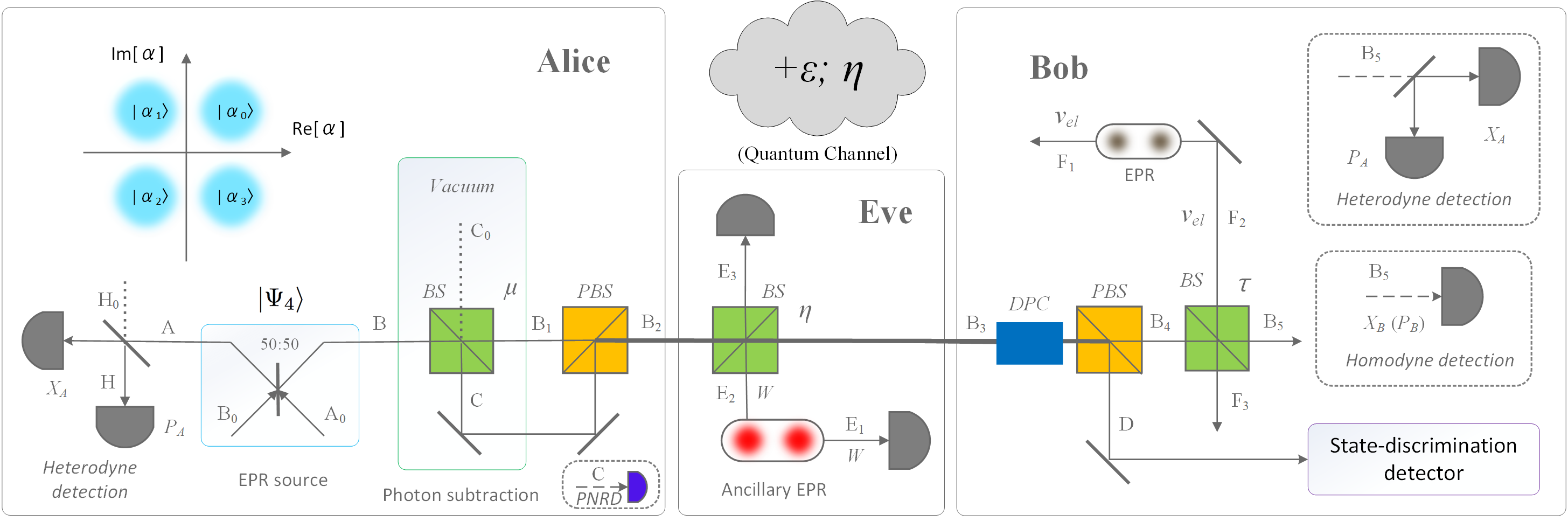}
\caption{Schematic diagram of the long-distance CVQKD. Alice detects one half of EPR state (the blue box) using heterodyne detection while another half is sent to the photon-subtraction module (the green box) which splits the incoming signal into two parts. The two parts are then recombined by using polarization-multiplexing technique and subsequently sent to Bob. Eve replaces the quantum channel and performs the optimal \emph{entangling cloner} attack during the transmission. Bob demultiplexes the incoming signal and measures one of them using homodyne or heterodyne detector, whereas the other mode is sent to the state-discrimination detector (the purple box). BS denotes beam splitter, PBS denotes polarizing beamsplitter, DPC denotes dynamic polarization controller, and PNRD stands for photon number resolving detector.}
\label{fig:LDSDR}
\end{figure*}

In what follows, we elaborate the long-distance discretely modulated CVQKD scheme. This novel scheme is based on the four-state CVQKD protocol so that its transmission distance could be extended more largely comparing with the continuous modulation counterparts. We focus on the principle of the whole long-distance CVQKD scheme first, leaving the detailed techniques description to the next section.

As shown in Fig. \ref{fig:LDSDR}, a source of the two-mode entangled state [Einstein-Podolsky-Rosen (EPR) state] is used for creating a secure key \cite{PhysRevA.77.012337}. After Alice prepares the entangled state $|\Psi_4\rangle$, she performs heterodyne detection on one half (mode $A$) of the state and sends another half (mode $B$) to the photon-subtraction operation (the module within the green box). This non-Gaussian operation is modeled by a beam splitter (BS) with transmittance $\mu$ and a vacuum state $|0\rangle$ imports the unused port of the beam splitter. As a result, the incoming signal (mode $B$) is then divided into two parts by the photon-subtraction operation. There are two advantages for applying photon-subtraction operation. Firstly, putting a proper non-Gaussian operation at Alice's side has been proven to be beneficial for lengthening the maximal transmission distance of the traditional CVQKD \cite{PhysRevA.87.012317}, because  this operation can be deemed the preparation trusted noise controlled by Alice, which can well prevent the eavesdropper from acquiring communication information \cite{Guo:2017ce,Usenko2016Trusted}. Secondly, the photon-subtraction operation tactfully provides a method to divide the incoming signal into two parts, which are the mode $B_1$ containing most photons for homodyne (or heterodyne) detector and the mode $C$ containing a few subtracted $j$ photons (or even one) for state-discrimination detector respectively. The two parts of signal are subsequently recombined by using the polarization-multiplexing technique with a polarizing beamsplitter (PBS). The recombinational mode $B_2$ is then sent to the lossy and insecure quantum channel.

In the EB CVQKD scheme, the quantum channel is replaced by an eavesdropper (say Eve) who performs the collective Gaussian attack strategy. This attacks is proved to be an optimal attack strategy in direct and reverse reconciliation protocols. Very recently, Leverrier \cite{PhysRevLett.118.200501} shows that it is sufficient to prove the security of CVQKD against collective Gaussian attacks in order to obtain security against general attacks, therefore confirming rigorously the belief that collective Gaussian attacks are indeed optimal against CVQKD. In this kind of attacks, Eve usually prepares her ancillary system in a product state and each ancilla interacts individually with a single pulse sent by Alice, being later stored in a quantum memory \cite{Garc2007Quantum}. The tripartite state then reads,
\begin{equation}
\begin{aligned}
&\rho_{ABE} =\big[\sum_{a}P(a)|a\rangle\langle a|_a\otimes\psi_{BE}^{a}\big]^{\otimes n}.
\end{aligned}
\end{equation}
After eavesdropping the communication revealed by Alice and Bob in the data post-processing, Eve applies the optimal collective measurement on the ensemble of stored ancilla to steal the secret information. In particular, Eve can launch the so called \emph{entangling cloner} \cite{PhysRevLett.97.190503,PhysRevLett.94.020505,PhysRevLett.101.200504} attack which is a kind of collective Gaussian attack. Specifically, Eve replaces the channel with transmittance $\eta$ and excess noise referred to the input $\chi$ by preparing the ancilla $|E\rangle$ with variance $W$ and a beam splitter with transmittance $\eta$. The value $W$ can be tuned to match the noise of the real channel $\chi_{line}=(1-\eta)/\eta+\varepsilon$. After that, Eve keeps one mode $E_{1}$ of $|E\rangle$ and injects the mode $E_{2}$ into the unused port of the beam splitter and thus acquires the output mode $E_{3}$. After repeating this process for each pulse, Eve stores her ancilla modes, $E_{1}$ and $E_{3}$, in quantum memories. Finally, Eve measures the exact quadrature on $E_{1}$ and $E_{3}$ after Alice and Bob reveal the classical communication information. The measurement of $E_{1}$ allows her to decrease the noise added by $E_{3}$.

After passing the untrusted quantum channel, Bob applies another PBS with dynamic polarization controller (DPC) to demultiplex the incoming signal. One of the demultiplexed modes $B_4$ is then sent to Bob's homodyne or heterodyne detector which is modeled by a BS with transmittance $\tau$ and its electronic noise is modeled by an EPR state with variance $v_{el}$. The mode $D$ is synchronously sent to the state-discrimination detector to improve the system's performance.

This long-distance CVQKD scheme subtly combines the merits of the four-state CVQKD protocol and photon-subtraction operation in terms of
lengthening maximal transmission distance,  surpassing the SQL via the state-discrimination detector.

\section{Techniques}

In this section we show the detailed characteristics of the photon-subtraction operation and the state-discrimination detector that can be used for beating the SQL.

\subsection{Photon-subtraction operation}

As shown in Fig. \ref{fig:LDSDR}, we suggest the EB CVQKD with photon-subtraction operation (the green box) applied at Alice's station, where other modules are temporarily ignored. Alice uses a beam splitter with transmittance $\mu$ to split the incoming mode $B$ and the vacuum state $C_{0}$ into modes $B_{1}$ and $C$. The yielded tripartite state $\rho_{ACB_{1}}$ can be expressed by
\begin{equation}
\begin{aligned}
\rho_{ACB_{1}} = U_{BS}[|\Psi\rangle_4\langle\Psi|_4\otimes|0\rangle\langle0|]U_{BS}^\dag.
\end{aligned}
\end{equation}
Subsequently a photon-number-resolving detector (PNRD, black dotted box at Alice's side) is adopted to measure mode $C$ by applying positive operator-valued measurement (POVM) $\{\hat{\Pi}_{0},\hat{\Pi}_{1}\}$ \cite{Eisaman2011Invited}. The photon number of subtraction $j$ depends on $\hat{\Pi}_{1}=|j\rangle\langle j|$. Only when the POVM element $\hat{\Pi}_{1}$ clicks can Alice and Bob keep $A$ and $B_{1}$. The photon-subtracted state $\rho_{AB_{1}}^{\hat{\Pi}_{1}}$ is given by
\begin{equation}
\begin{aligned}
\rho_{AB_{1}}^{\hat{\Pi}_{1}} =
\frac{\mathrm{tr}_{C}(\hat{\Pi}_{1}\rho_{ACB_{1}})}{\mathrm{tr}_{ACB_{1}}(\hat{\Pi}_{1}\rho_{ACB_{1}})}
\end{aligned},
\end{equation}
where $\mathrm{tr}_{X}(\cdot)$ is the partial trace of the multi-mode quantum state and $\mathrm{tr}_{ACB_{1}}(\hat{\Pi}_{1}\rho_{ACB_{1}})$ is the success probability of  subtracting $j$ photons, which can be calculated as
\begin{equation}
\begin{aligned}\label{7}
P^{\hat{\Pi}_{1}}_{(j)}& =
\mathrm{tr}_{ACB_{1}}(\hat{\Pi}_{1}\rho_{ACB_{1}})  \\
&=(1-\xi^2)\sum_{n=j}^{\infty}C_{n}^{j}\xi^{2n}(1-\mu)^{j}\mu^{n-j}\\
&=\frac{(1-\xi^2)(1-\mu)^{j}\xi^{2j}}{(1-\mu\xi^{2})^{j+1}},
\end{aligned}
\end{equation}
where $C_{n}^{j}$ is combinatorial number and $\xi=\frac{\alpha}{\sqrt{1+\alpha^2}}$. 


After passing the BS, it is worth noticing that the subtracted state $\rho_{AB_{1}}^{\hat{\Pi}_{1}}$ is not Gaussian anymore, while its entanglement degree increases with the introduction of the photon-subtraction operation \cite{Guo:2017ce,PhysRevA.87.012317}.

Due to the fact that heterodyne detection on one half of the EPR state will project the other half onto a coherent state, which is convenient to implement in experimentation, we take into account a situation where Alice performs heterodyne detection and Bob executes homodyne detection.
Suppose $\Gamma_{AB_{1}}^{(j)}$ represents the covariance matrix of $\rho_{AB_{1}}^{\hat{\Pi}_{1}}$, and it can be given by
\begin{equation}\label{9}       
\Gamma_{AB_{1}}^{(j)}=
\left(                 
\begin{array}{cc}   
  X'\mathbb{I} & Z_4'\sigma_{z}\\  
    Z_4'\sigma_{z} & Y'\mathbb{I}\\  
  \end{array}
\right),                 
\end{equation}
where 
\begin{equation}\label{qw}
\begin{aligned}
Z_4'&=\frac{\sqrt{\mu}\xi(j+1)}{1-\mu\xi^{2}},\\
X'&=\frac{\mu\xi^{2}+2j+1}{1-\mu\xi^{2}},\\
Y'&=\frac{\mu\xi^{2}(2j+1)+1}{1-\mu\xi^{2}}.
\end{aligned}
\end{equation}
See \cite{PhysRevA.93.012310} for the detailed calculations.

Note that for the proposed long-distance CVQKD scheme, the PNRD which is placed at Alice's side is removed, whereas the subtracted mode $C$ which is supposed to enter the PNRD is recombined with mode $B_1$ in a PBS by using polarization-multiplexing technique. The task of resolving subtracted photon number is therefore handed over to the state-discrimination detector at Bob's side.

\subsection{State-discrimination detector}

We design a state-discrimination detector to increase the performance of the CVQKD coupled with photon-subtraction operation. This quantum detector can unconditionally discriminate four nonorthogonal coherent states in QPSK modulation with the error probabilities lower than the SQL.

\begin{figure}
\centering
\includegraphics[width=3.2in,height=2.1in]{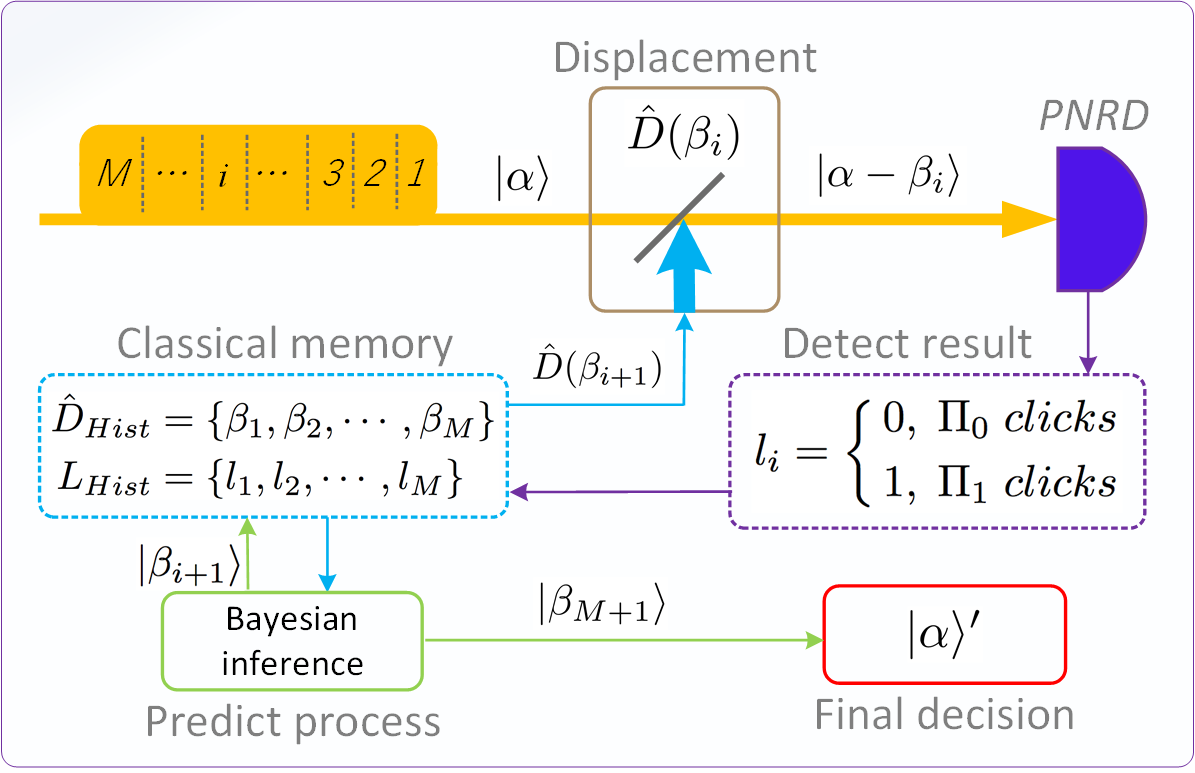}
\caption{Schematic diagram of the improved state-discrimination detector using photon number resolving and adaptive measurements in the form of feedback.}
\label{fig:SDR}
\end{figure}

As shown in Fig. \ref{fig:SDR}, we depict the structure of the improved state-discrimination detector using photon number resolving and adaptive measurements \cite{Higgins:2007ig,Armen:2002kf,Wiseman:1995ch} in the form of fast feedback. This state-discrimination detector contains $M$ times adaptive measurements in the field of $|\alpha\rangle$. For each measurement $i\,(i\in\{0,1,\cdots,M\})$, the strategy first prepares a predicted state $|\beta_i\rangle$ which has the highest probability based on the current data in classical memory. Subsequently, a displacement $\hat{D}(\beta_i)$ is adopted to displace $|\alpha\rangle$ to $|\alpha-\beta_i\rangle$ and a PNRD is used to detect the number of photons of the displaced field. If the predicted state is correct, i.e., $|\beta_i\rangle=|\alpha\rangle$, $\Pi_0$ will click, because the input field is displaced to vacuum so that the PNRD cannot detect any photon \cite{Becerra:2013jw}. Note that different from the photon-subtraction operation where $\Pi_0$ clicks represents the failure of subtracting photon, $\Pi_0$ clicks here denotes that the improved state-discrimination strategy has correctly predicted the input state. This successful prediction is marked as $l_i=0$, otherwise $l_i=1$. After the $i$-th adaptive measurement, the strategy calculates the posterior probabilities of all possible states ($|\alpha_{i0}\rangle$, $|\alpha_{i1}\rangle$,$|\alpha_{i2}\rangle$ and $|\alpha_{i3}\rangle$) using Bayesian inference according to the present label history $L_{Hist}$ and predicted history $\hat{D}_{Hist}$ (Note that for now $\beta_i$ has aready been added to the $\hat{D}_{Hist}$ with previous data to collectively calculate these probabilities), and designates the most probable state as $|\beta_{i+1}\rangle$, which is deemed as an input for next feedback. In each feedback period, the probabilities of all possible states are updated dynamically and the posterior probabilities of period $i$ become prior probabilities in period $i+1$. The rule of Bayesian inference can be expressed as
\begin{equation} \label{popr}
\begin{aligned}
\bm{P}_{\mathrm{po}}(\{|\alpha\rangle\}|\beta_i,l_i)
=
A\,
\mathbb{P}(l_i|\beta_i,\{|\alpha\rangle\})
\bm{P}_{\mathrm{pr}}(\{|\alpha\rangle\}),
\end{aligned}
\end{equation}
where $\bm{P}_{\mathrm{po}}(\{|\alpha\rangle\}|\beta_i,l_i)$ and $\bm{P}_{\mathrm{pr}}(\{|\alpha\rangle\})$ are the posterior and prior probabilities respectively, $\mathbb{P}(l_i|\beta_i,\{|\alpha\rangle\})$ is conditional Poissonian probability of observing the detection result $l_i$ for $|\alpha\rangle$ displaced by field $\beta_i$, and $A$ is the normalization factor calculated by summing Eq. (\ref{popr}) over all possible states. Therefore, the final decision $|\beta_{M+1}\rangle$ of the input state $|\alpha\rangle$ can be predicted in the last adaptive measurement $M$ using iterative Bayesian inference \cite{Becerra:2011hb}.

This kind of strategies could surpass the SQL and approach the Helstrom bound with the help of high bandwidth and high detection efficiency. Mathematically, the SQL for discriminating the four nonorthogonal coherent state in QPSK modulation can be expressed by
\begin{equation}
\begin{aligned}
P_{SQL}=1-\left[1-\frac{1}{2}\mathrm{erfc}\bigg(\sqrt{\frac{|\alpha|^2}{2}}\bigg) \right]^2,
\end{aligned}
\end{equation}
where 
\begin{equation}
\begin{aligned}
\mathrm{erfc}(x)=\frac{2}{\sqrt{\pi}}\int_x^\infty e^{-t^2}\mathrm{d}t, 
\end{aligned}
\end{equation}
and the Helstrom bound for the QPSK signals can be approximated by using the square-root measure (SRM) \cite{PhysRevA.89.042330}, which can be calculated by
\begin{equation}
\begin{aligned}
P_{Hel}=1-\frac{1}{16}\bigg(\sum_{k=1}^4\sqrt{\omega_k}\bigg)^2,
\end{aligned}
\end{equation}
where $\omega_k=e^{-\alpha^2}\sum_{n=1}^{4}\mathrm{exp}{[(1-k)\frac{2\pi in}{4}+\alpha^2\mathrm{exp}({\frac{2\pi in}{4}})]}$ are eigenvalues of Gram matrix for QPSK signals. As the improved state-discrimination detector is parallel with homodyne or heterodyne detector at Bob's side, the ultimate detection of the states is codetermined by the coherent detector and the state-discrimination detector. From the perspective of information-theoretical sense, we can define an improvement ratio $\zeta$ to depict how much performance could the state-discrimination detector enhance the CVQKD system, namely
\begin{equation}
\begin{aligned}
\zeta=\frac{1-P_{rec}^{(M)}}{1-P_{SQL}},
\end{aligned}
\end{equation}
where $P_{rec}^{(M)}$ represents the error probability of state-discrimination detector with $M$ adaptive measurements. Theoretically, the detector could reach the Helstrom bound when $M$ is large enough. Therefore, the optimal improvement ratio $\zeta_{opt}$ can be calculated by considering the minimum error probability allowed by quantum mechanics. Thus we have
\begin{equation}
\begin{aligned}
\zeta_{opt}=\frac{1-P_{Hel}}{1-P_{SQL}}.
\end{aligned}
\end{equation}
\begin{figure}
\centering
\includegraphics[width=3.4in,height=2.6in]{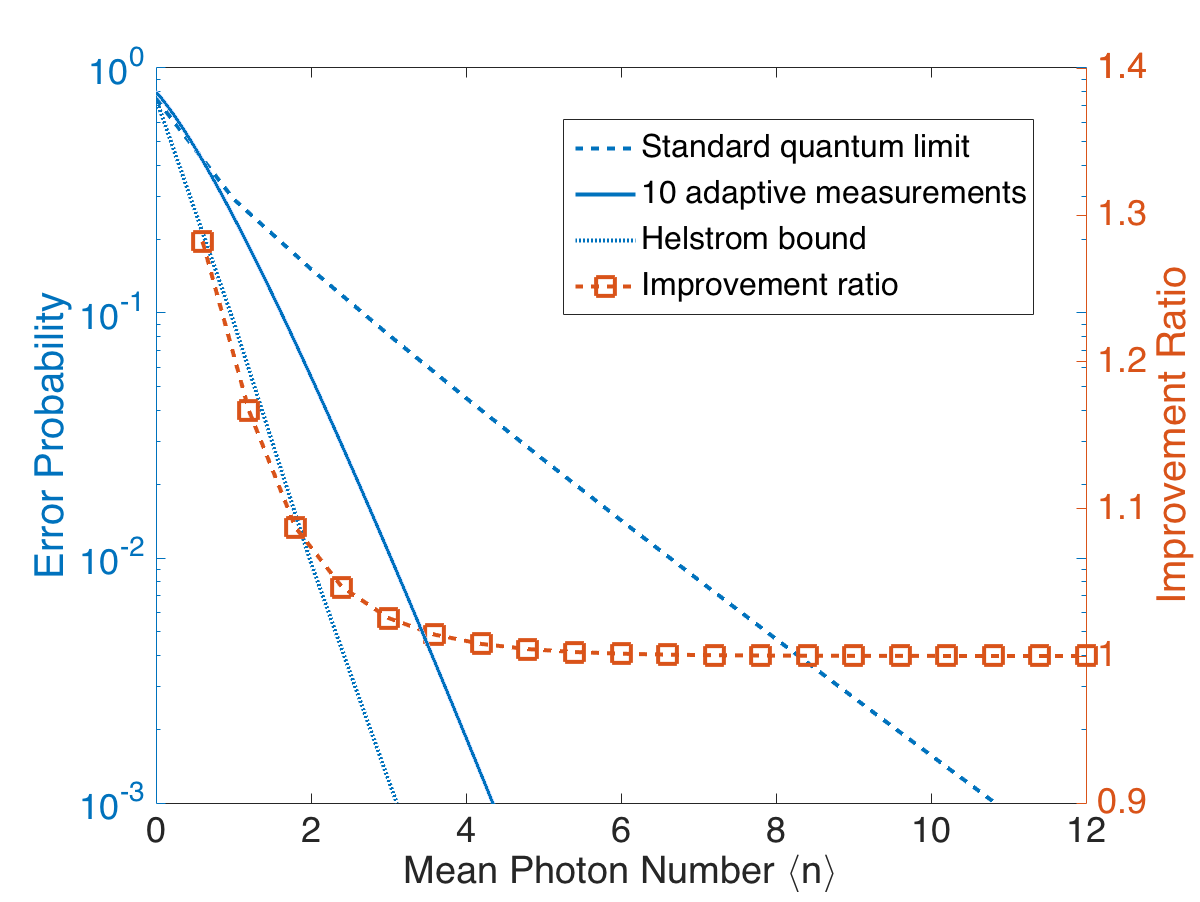}
\caption{Error probabilities for discriminating QPSK states and improvement ratio as functions of mean photon number $\langle n\rangle$. The dashed line denotes standard quantum limit, the solid line denotes the state-discrimination detector with 10 adaptive measurements, the dotted line denotes Helstrom bound and the red dashed line with squares denotes improvement ratio.}
\label{fig:EP}
\end{figure}

In Fig. (\ref{fig:EP}), we illustrate the error probabilities of discriminating the four nonorthogonal coherent states and the improvement ratio as functions of mean photon number $\langle n\rangle$. The blue dashed line shows the standard quantum limit to which the conventional and ideal coherent detector can achieve, while the blue solid line shows that the state-discrimination detector with $10$ adaptive measurements is below the SQL and approaches the Helstrom bound (the blue dotted line). The red dashed line with squares denotes the optimal improvement ratio which descends quickly with the increased mean photon number but still above $1$. Therefore, the proposed detection strategy that consists of state-discrimination detector and coherent detector could improve the performance of CVQKD system, satisfying the requirement of long-distance transmission.  

\section{Performance and discussion}
In this section, we show the performance of the proposed long-distance CVQKD scheme with numeric simulation results. To simplify the expression, we only focus on a scenario that Bob performs homodyne detection and reverse reconciliation (RR) in the data post-processing procedure. 

\subsection{Parameter optimization}

\begin{figure}
\centering
\includegraphics[width=3.6in,height=2.9in]{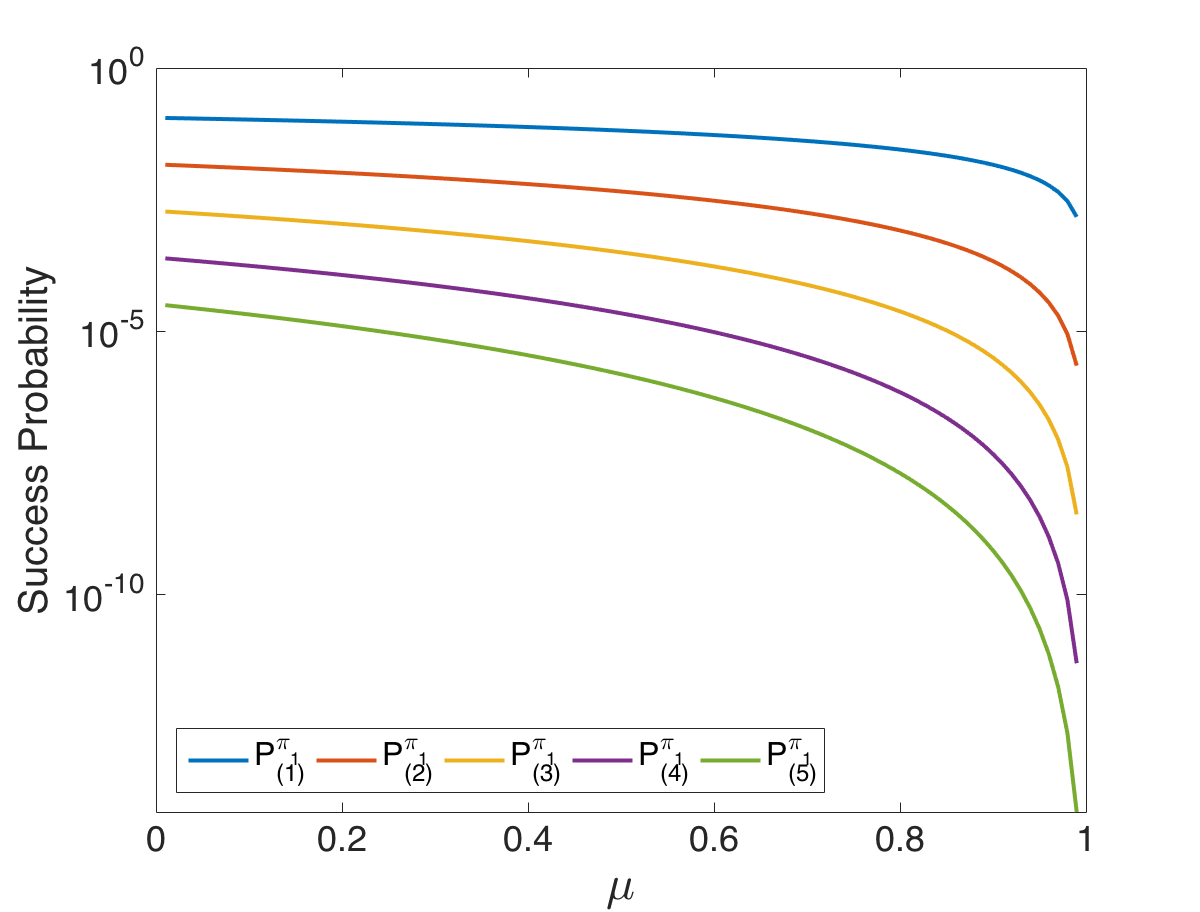}
\caption{Success probability of subtracting $j$ photons for discretely modulated CVQKD with different transmittances $\mu$. The lines from top to bottom represent one-photon subtraction (blue line), two-photon subtraction (red line), three-photon subtraction (yellow line), four-photon subtraction (purple line) and five-photon subtraction (green line).}
\label{fig:successProbability}
\end{figure}

We first demonstrate the optimal values of simulated parameters before giving the performance of secret key rate. It is known that the optimal photon-subtraction operation in Gaussian-modulated CVQKD can be achieved when only one photon is subtracted \cite{Guo:2017ce,PhysRevA.93.012310}, which means that subtracting one photon is the preferred operation to improve the transmission distance. For the proposed long-distance discretely-modulated CVQKD scheme,  we show the success probability of subtracting $j\,(j=1,2,3,4,5)$ photons as a function of transmittance $\mu$ in Fig. (\ref{fig:successProbability}). Similar to its Gaussian-modulated counterpart, the success probability of subtracting one photon ($j=1$, blue line) outperforms other numbers of photon subtraction and the success probability decreases with the increase of the number of subtracted photons. Meanwhile, as shown in Fig. (\ref{fig:EP}), the red dashed line with squares depicts the improvement ratio of the improved state-discrimination detector. It is obvious that the highest value of the improvement can be obtained with mean photon number $\langle \bar{n}\rangle=1$. This coincidence ($j=\langle \bar{n}\rangle=1$) allows us to obtain the optimal performance by tactfully combining photon-subtraction operation and state-discrimination detector together. More specifically, the one photon subtracted by photon-subtraction operation at Alice' side is detected by state-discrimination detector at Bob' side, and both modules perform optimally as one photon meets the optimal requirements. Therefore, we consider the optimal one-photon subtraction operation in subsequent simulations to show the best performance of the proposed scheme.

\begin{figure}
\centering
\includegraphics[width=3.6in,height=2.9in]{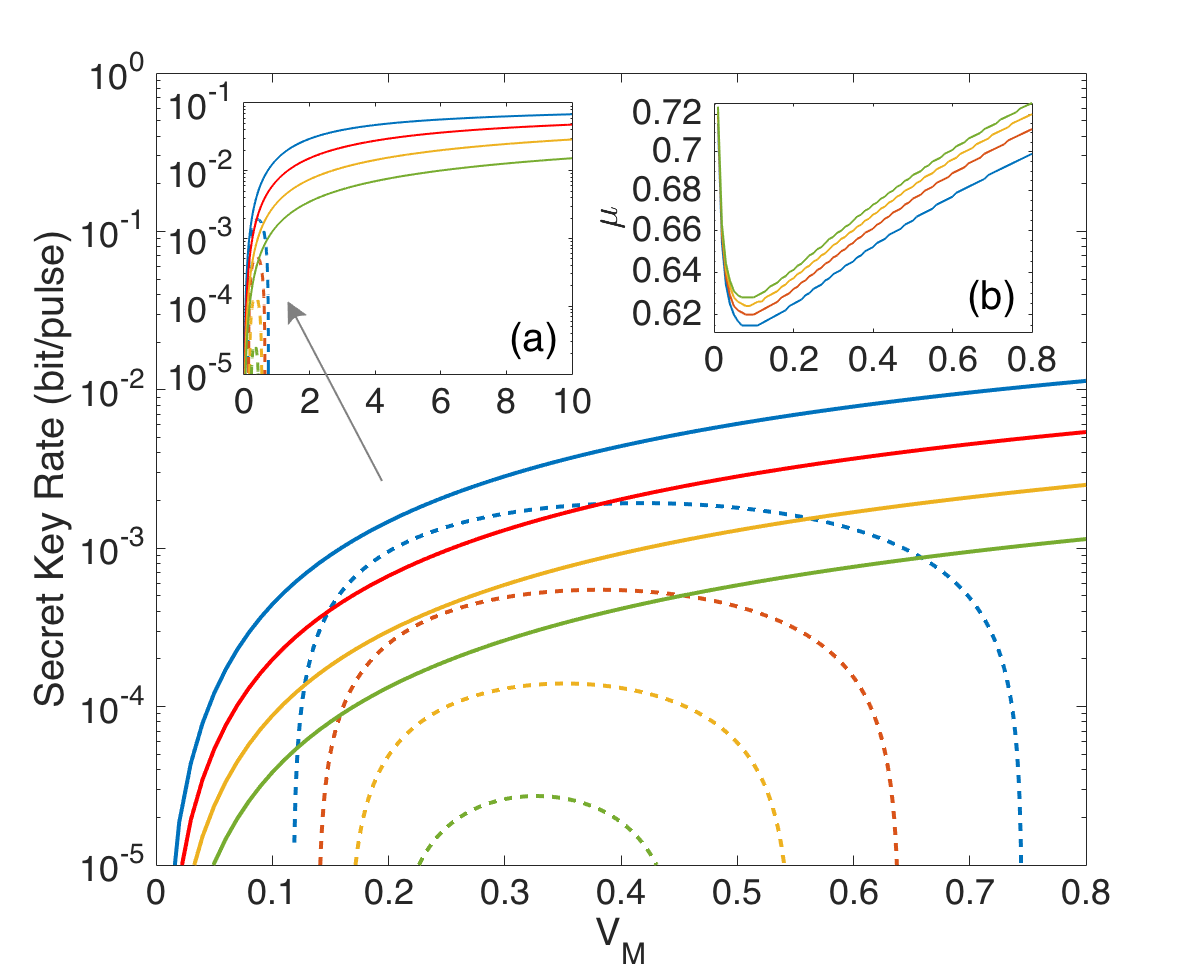}
\caption{Asymptotic secret key rate as a function of modulation variance $V_M$ in different channel losses with excess noise $\varepsilon=0.01$. Solid lines denote the proposed long-distance CVQKD scheme with optimal one-photon subtraction, dashed lines represent the four-state CVQKD protocol. Channel losses are set to $12$ dB (blue lines), $16$ dB (red lines), $20$ dB (yellow lines) and $24$ dB (green lines), respectively. Inset (a) is the extended graph with $V_M$ to $10$. Inset (b) shows the optimal $\mu$ for the current secret key rate as a function of modulation variance $V_M$.}
\label{fig:compareDiffVA}
\end{figure}
\begin{figure}
\centering
\includegraphics[width=3.6in,height=2.9in]{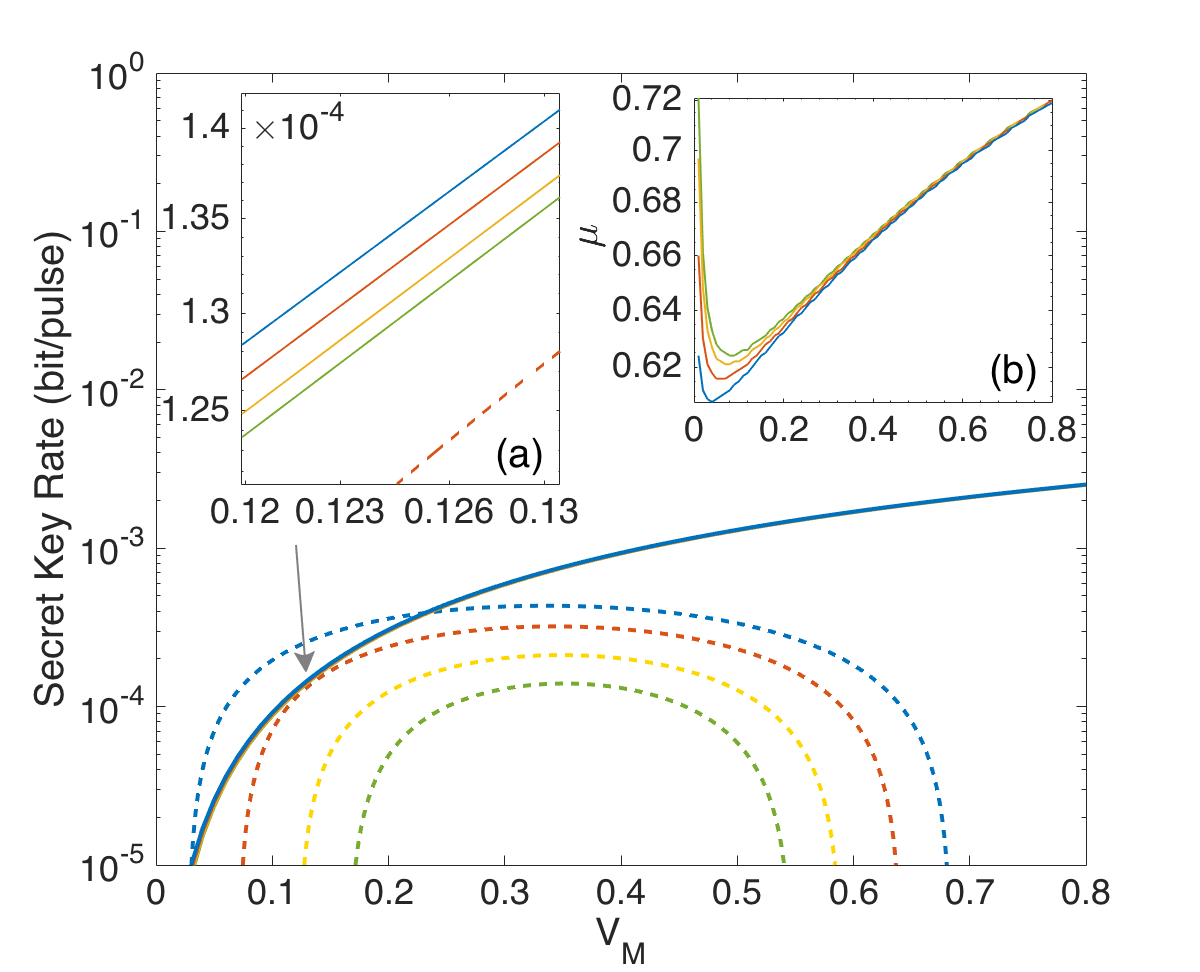}
\caption{Asymptotic secret key rate as a function of modulation variance $V_M$ in different excess noises with transmission distance $d=100$ km. Solid lines denote the proposed long-distance CVQKD scheme with optimal one-photon subtraction, dashed lines represent the four-state CVQKD protocol. Excess noises are set to 0.002 (blue lines), 0.005 (red lines), 0.008 (yellow lines) and 0.01 (green lines), respectively. Inset (a) is the magnified graph with $V_M$ limited from 0.12 to 0.13. Inset (b) shows the optimal $\mu$ for the current secret key rate as a function of modulation variance $V_M$.}
\label{fig:compareDiffVAepsilon}
\end{figure}

Because channel loss and excess noise are two of the most important factors that would have an effect on the performance of CVQKD system \cite{Fossier:2009dz}, the performance of these parameters with different modulation variance $V_M$ needs to be illustrated. In Fig. (\ref{fig:compareDiffVA}) and Fig. (\ref{fig:compareDiffVAepsilon}), solid lines denote the performance of the proposed long-distance CVQKD scheme with the optimal one-photon subtraction operation, while dashed lines represent the four-state CVQKD protocol as a comparison, and their secret key rates change as $V_M$ changes. The global simulation parameters are as follows: reconciliation efficiency is $\beta=95\%$, quantum efficiency of Bob's detection is $\tau=0.6$ and electronic noise is $v_{el}=0.05$. In Fig. (\ref{fig:compareDiffVA}), excess noise $\varepsilon$ and other parameters are fixed to legitimate values, the numerical areas of $V_M$ are compressed for the four-state CVQKD protocol when channel loss increases, and the secret key rate decreases rapidly with the increase of channel loss. While for the proposed long-distance CVQKD scheme, $V_M$ can be set to a large range of values and its secret key rate increases with the increased $V_M$ even though the secret key rate also decreases as channel loss increases, which means the performance of the proposed long-distance CVQKD scheme would be consecutively improved theoretically when the modulation variance is set large enough. However, this cannot be realized in practice, thus the modulation variance $V_M$ must be set to a reasonable value in simulations. In Fig. (\ref{fig:compareDiffVAepsilon}), transmission distance, which is proportional to the channel loss (0.2 dB/km), and other parameters are fixed. For the four-state CVQKD protocol, its optimal regions of $V_M$ are also compressed with the  increased excess noise $\varepsilon$. Fortunately, there is only slight impact on the proposed long-distance CVQKD scheme with one-photon subtraction when excess noise $\varepsilon$ changes. It shows the proposed scheme greatly outperforms the four-state CVQKD protocol in terms of tolerable channel excess noise. The reasons may be given as follows. Firstly, excess noise can be deemed channel imperfections which deteriorate the correlation between the two output modes, while photon-subtraction operation can well enhance the correlation which is positively related to entanglement degree of EPR state and thus improves the performance of CVQKD system \cite{Guo:2017ce,PhysRevA.87.012317}. Rendering the CVQKD system that applied this non-Gaussian operation tolerates more higher excess noise. Secondly, the proposed long-distance CVQKD scheme is not very sensitive to the noise with the help of state-discrimination detector, which means Bob can obtain more correct results without performing very precise measurement on quadrature $\hat{x}$ or/and $\hat{p}$. The reason is that the raw key in Gaussian modulated CVQKD protocol is tremendously affected by channel excess noise (and imperfect coherent detector) since its information is directly encoded in quadratures. In the four-state CVQKD protocol, the information is encoded in QPSK modulation which can be unconditionally discriminated by the state-discrimination detector \cite{Becerra:2013jw}. Therefore, the detection strategy could predict incoming state using probability-based method, i.e. Bayesian inference, thus alleviating the impact of excess noise.

\subsection{Secret key rates}

Up to now, we have derived the parameters that may largely affect the CVQKD system. In what follows, we consider the secret key rate of the proposed CVQKD scheme. In general, the asymptotic secret key rate can be calculated with the form
\begin{equation}\label{k}
\begin{aligned}
K_{asym}=\beta I(A:B)-S(E:B),
\end{aligned}
\end{equation}
where $\beta$ is the efficiency for RR, $I(A:B)$ is the Shannon mutual information between Alice and Bob, and $S(E:B)$ is the Holevo bound \cite{Nielsen2011Quantum} of the mutual information between Eve and Bob. For the proposed CVQKD protocol, the asymptotic secret key rate in Eq. (\ref{k}) can be rewritten by
\begin{equation}
\begin{aligned}
K_{asym}=P^{\hat{\Pi}_{1}}_{(j)}[\beta \zeta_{opt} I(A:B)-S(E:B)].
\end{aligned}
\end{equation}
As previously mentioned, $P^{\hat{\Pi}_{1}}_{(j)}$ represents the probability of successful subtracting $j$ photons and $\zeta_{opt}$ depicts the improvement ratio of the introduced state-discrimination detector. Detailed calculation of the asymptotic secret key rate can be found in Appendix A. 

In Fig. (\ref{fig:CP}), we depict the asymptotic secret key rate as a function of transmission distance for the CVQKD protocol. Red line shows the original four-state protocol proposed in \cite{Leverrier:2009cr}, yellow line denotes the optimal one-photon subtraction scheme for Gaussian modulated coherent state proposed in \cite{PhysRevA.93.012310}, blue line represents the scheme of four-state protocol with one-photon subtraction, and the green line denotes the proposed long-distance CVQKD scheme using non-Gaussian state-discrimination detector. The modulation variance $V_M$ of above protocols is optimized except for the proposed long-distance CVQKD scheme since its secret key rate is monotonic increasing in a large range of the modulation variance $V_M$, which means that the performance of the proposed scheme can be further improved when $V_M$ is set to larger value. However, from the perspective of fair comparison and practical significance, the modulation variance $V_M$ of the proposed long-distance CVQKD scheme is reasonably set as same as its fundamental communication protocol, i.e., the four-state CVQKD protocol. As shown in Fig. (\ref{fig:CP}), the proposed long-distance CVQKD scheme outperforms all other CVQKD protocols in terms of maximum transmission distance up to 330 km. Therefore, the proposed long-distance CVQKD scheme using non-Gaussian state-discrimination detector could be more suitable for long-distance transmission. Note that this distance record is limited by the secret key rate more than $10^{-6}$ bits per pulse, and it can be further extended when one considers the secret key rate below this bound.

\begin{figure}
\centering
\includegraphics[width=3.6in,height=2.9in]{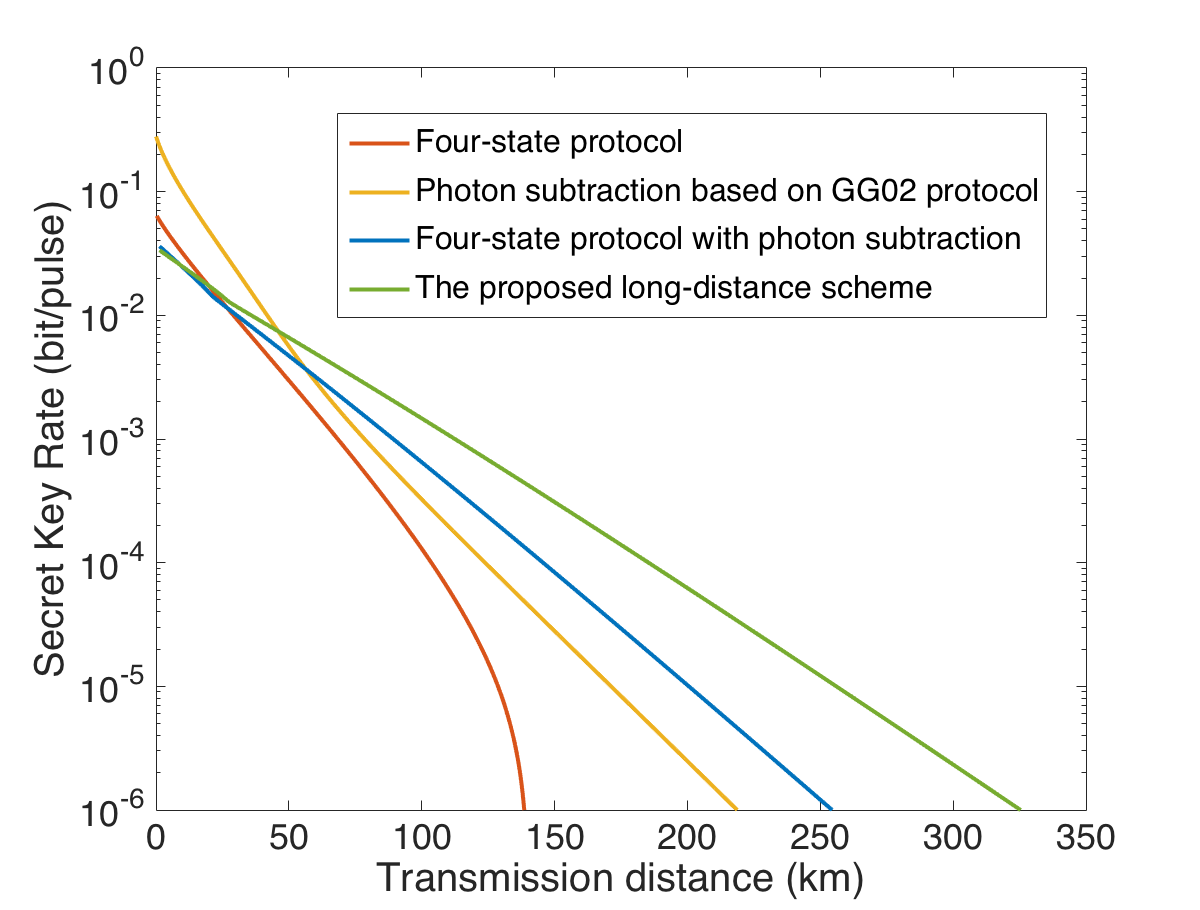}
\caption{Asymptotic secret key rate as a function of transmission distance with excess noise $\varepsilon=0.01$. Red line shows the original four-state protocol, yellow line denotes the optimal one-photon subtraction scheme for Gaussian modulated coherent state, blue line represents the scheme of four-state protocol with one-photon subtraction, and the green line denotes the proposed long-distance scheme using non-Gaussian state-discrimination detector.}
\label{fig:CP}
\end{figure}

\begin{figure}
\centering
\includegraphics[width=3.6in,height=2.9in]{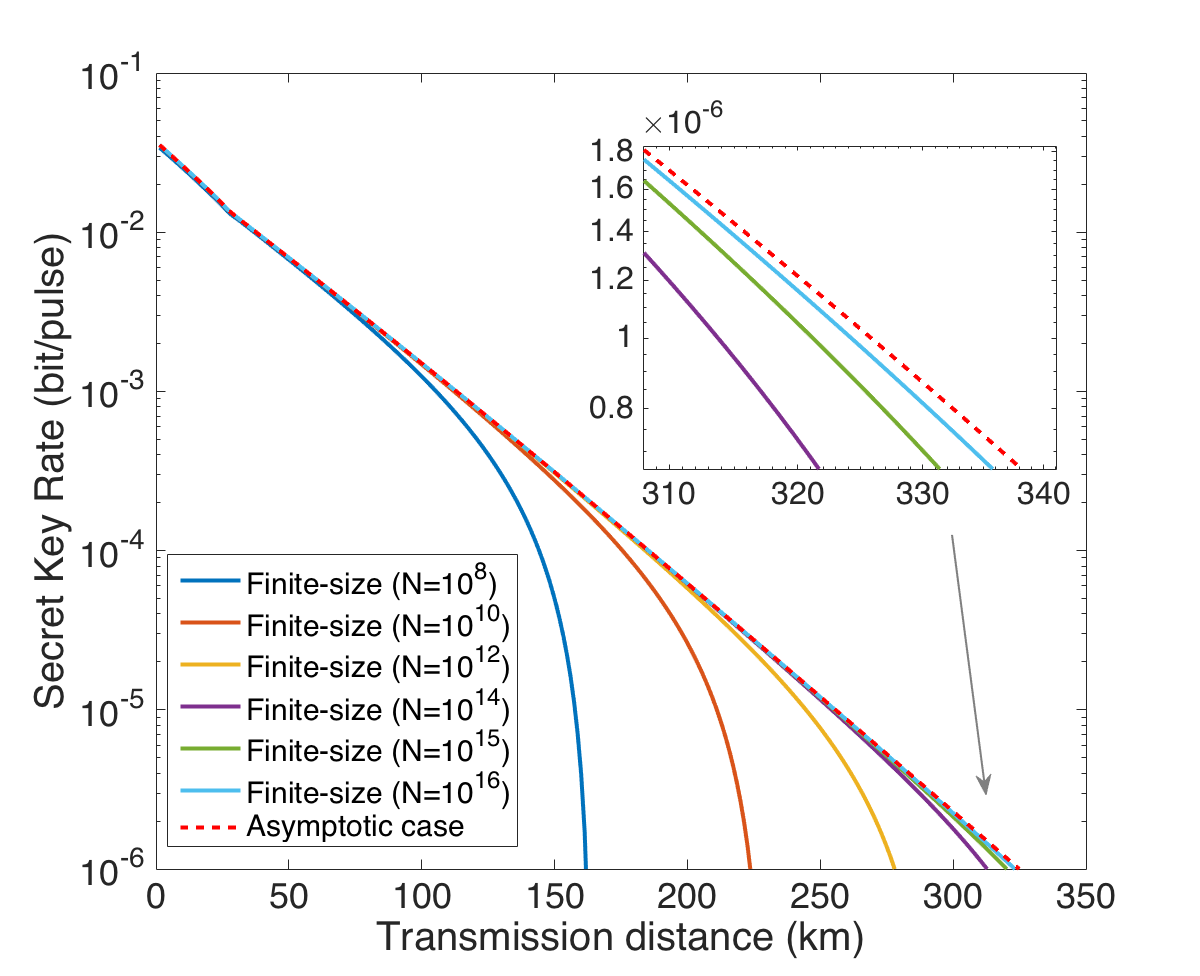}
\caption{Finite-size secret key rate of the proposed long-distance CVQKD scheme with one-photon subtraction as a function of transmission distance. The excess noise is set to $\varepsilon=0.01$. From left to right, the solid lines correspond to block lengths of $N=10^8,10^{10},10^{12},10^{14},10^{15}$ and $10^{16}$,  and  the red dashed line denotes the asymptotic case. The secret key rate is null for a block length of $10^4$.}
\label{fig:finite}
\end{figure}

In addition, finite-size effect \cite{Leverrier:2010es} needs to be taken into consideration, since the length of secret key is impossibly unlimited in practice. Moreover, one can make the assumption in the asymptotic case that the quantum channel is perfectly known before the transmission is performed, while in finite-size scenario, one actually does not know the characteristics of the quantum channel in advance. Because a part of exchanged signals has to be used for parameter estimation rather than generates the secret key. As shown in Fig. (\ref{fig:finite}), the performance of the proposed CVQKD scheme in finite-size regime is outperformed by that obtained in asymptotic limit. The maximum transmission distance significantly decreases when the number of total exchanged signals $N$ decreases. However, it still has a large improvement when comparing with original four-state CVQKD protocol and its Gaussian-modulated protocol counterpart which also take finite-size effect into account. Notice that the performance in the finite-size regime will converge to the asymtotic case if $N$ is large enough. The detailed calculation of secret key rate in the finite-size regime can be found in Appendix B.

\begin{figure}
\centering
\includegraphics[width=3.6in,height=2.85in]{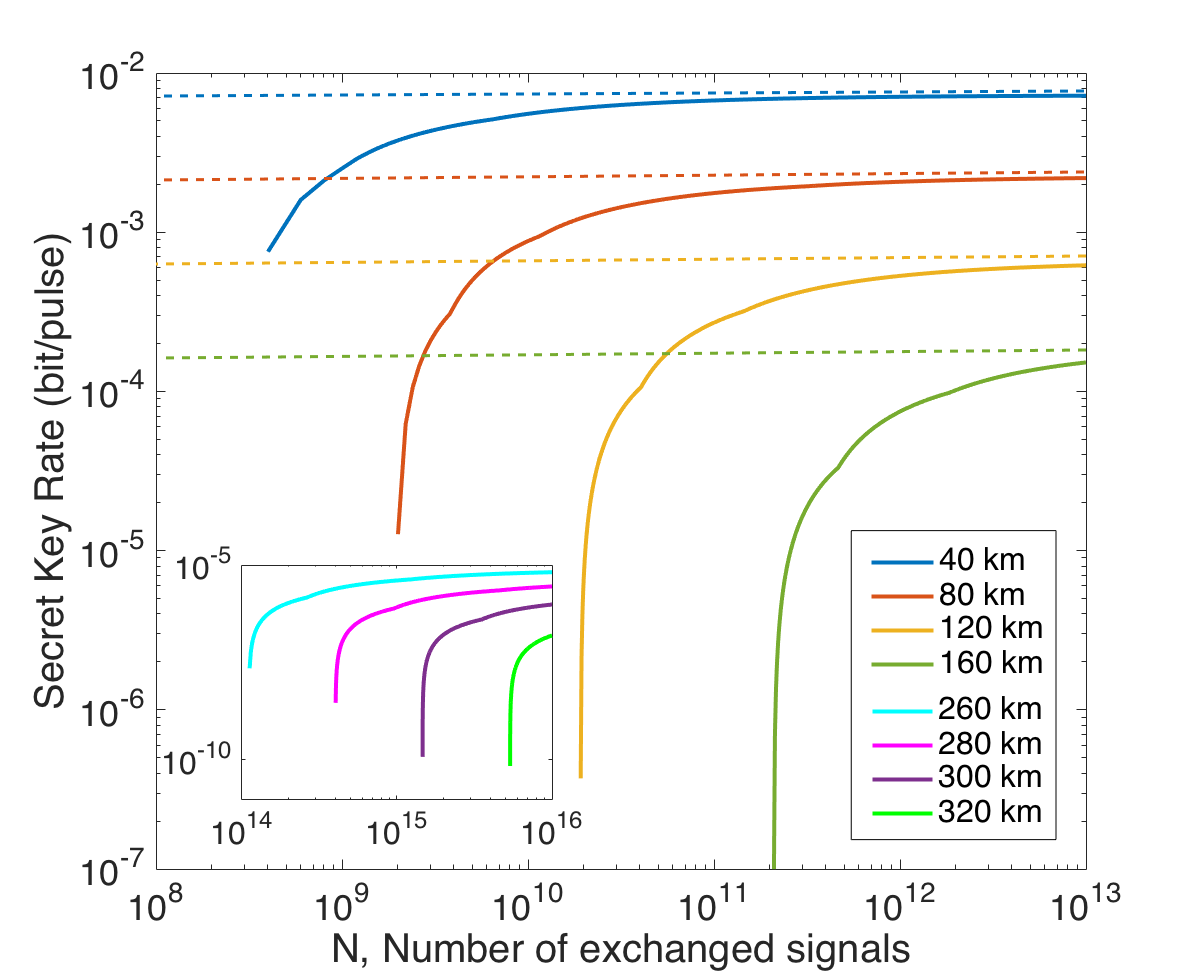}
\caption{Composable secret key rate of the proposed long-distance CVQKD with one-photon subtraction as a function of $N$, the number of exchanged signals. From top to bottom, solid lines denote the distances of $d=40,80,120$ and $160$ km. The dashed lines correspond to the respective asymptotic case. Inset shows the composable secret key rate of the proposed scheme at long-distance range, the lines from top to bottom denote the distances of $d=260,280,300$ and $320$ km, respectively. Excess noise is $\varepsilon=0.01$, discretization parameter is $d=5$, robustness parameter is $\epsilon_{rob}\le 10^{-2}$ and security parameter is $\epsilon=10^{-20}$. Other intermediate parameters can be found in Appendix C.}
\label{fig:composable}
\end{figure}

Finally, we demonstrate the performance of the proposed long-distance CVQKD scheme in composable security framework. The composable security is the enhancement of security based on uncertainty of the finite-size effect \cite{PhysRevLett.109.100502} so that one can obtain the tightest secure bound of the protocol by carefully considering every detailed step in CVQKD system \cite{Leverrier:2015he}. In Fig. (\ref{fig:composable}), we show the secret key rate of the proposed long-distance CVQKD scheme with one-photon subtraction operation in the case of composable security, as a function of total exchanged signals $N$. The performance is more pessimistic than that obtained in the finite-size regime, let alone in the asymptotic limit. For example, assuming that $N=10^{14}$ and the minimal secret key rate is limited to above $10^{-6}$ bis per pulse, the maximal transmission distance in finite-size regime is approximate $320$ km (purple line in Fig. (\ref{fig:finite})), while the maximal transmission distance is reduced to approximate $260$ km (light blue line in Fig. (\ref{fig:composable})) when one considers the proposed scheme in composable security framework. Therefore, the composable security, which takes the failure probabilities of every step into account, is the strictest theoretic security analysis of CVQKD system so that one can obtain more practical secure bound. In addition, the composable secret key rate also approaches the asymptotic value for very large $N$ (dashed lines). The detailed calculation of the secret key rate for composable security is shown in Appendix C.

The asymptotic limit, the finite-size scenario and the composable security framework are the efficient approaches to evaluate the performance of CVQKD system. Although the results vary with the different approach, the trends of the performance are similar. Therefore, the proposed long-distance CVQKD using non-Gaussian state-discrimination detector can beat other existing CVQKD protocols in terms of maximal transmission distance and thus meet the requirement of long-distance transmission.

\section{Conclusion}
We have suggested a novel long-distance CVQKD using non-Gaussian state-discrimination detector. The discretely-modulated four-state CVQKD protocol is adopted as the fundamental communication protocol since it can well tolerate the lower SNR and hence it is more suitable for the long-distance transmission compared with Gaussian-modulated counterpart. We deploy a non-Gaussian operation, i.e. photon-subtraction operation at the transmitter, where the photon-subtraction operation is not only used for splitting the signal, but also used for lengthening the transmission distance of CVQKD. Meanwhile, an improved state-discrimination detector is applied at the receiver to codetermine the measurement result with coherent detector. The state-discrimination detector can be deemed as the optimized quantum measurement for the received nonorthogonal coherent states, beating the standard quantum limit using adaptive measurements in the form of fast feedback. Therefore, Bob can obtain more precise result of incoming signal in the QPSK modulation with the help of the state-discrimination detector. By exploiting multiplexing technique, the yielded signals are simultaneously transmitted  through an untrusted quantum channel, and subsequently sent to the state-discrimination detector and coherent detector, respectively. Security analysis shows that the proposed scheme can lengthen the maximal transmission distance, and thus outperform other existing CVQKD protocols. Furthermore, by taking the finite-size effect and the composable security into account we obtain the tightest bound of the secure distance, which is more practical than that obtained in asymptotic limit. In terms of possible future research, it would be interesting to design an experiment to implement this long-distance CVQKD scheme for its practical security analysis.

\begin{acknowledgments}
This work is supported by the National Natural Science Foundation of China (Grant No. 61379153, No. 61572529), and the Fundamental Research Funds for the Central Universities of Central South University (Grant No. 2017zzts147).
\end{acknowledgments}

\begin{appendix}

\section{Calculation of asymptotic secret key rate}

We consider the calculation of asymptotic secret key rates of the proposed long-distance CVQKD scheme where Alice performs heterodyne detection and Bob performs homodyne detection respectively. Note that the state $\rho_{AB_{1}}^{\hat{\Pi}_{1}}$ is not Gaussian anymore after photon-subtraction operation, we thus cannot directly use results of the conventional Gaussian CVQKD to calculate the secret key rate. Fortunately, the secret key rate of state $\rho_{AB_{1}}^{\hat{\Pi}_{1}}$ is more than that of the Gaussian state $\rho_{AB_{1}}^{\hat{G}}$ counterpart which has the identical covariance matrix according to extremity of the Gaussian quantum states \cite{PhysRevLett.97.190503,PhysRevLett.97.190502,PhysRevLett.96.080502}. Therefore, the lower bound of the asymptotic secret key rate under optimal collective attack can be given by
\begin{equation}
\begin{aligned}
K_{asym}=P^{\hat{\Pi}_{1}}_{(j)}[\beta \zeta_{opt} I(A:B)-S(E:B)],
\end{aligned}
\end{equation}
where $\beta$ is the efficiency for reverse reconciliation, $I(A:B)$ is the Shannon mutual information between Alice and Bob, and $S(E:B)$ is the Holevo bound \cite{Nielsen2011Quantum} of the mutual information between Eve and Bob. 

Assuming that Alice's heterodyne detection and PBSs used for multiplexing are perfect, and Bob's homodyne detector is characterized by an transmittance $\tau$ and electronic noise $v_{el}$, then the detection-added noise referred to Bob's input can be given by $\chi_{hom}=[(1-\tau)+v_{el}]/\tau$. In addition, the channel-added noise is expressed by $\chi_{line}=(1-\eta)/\eta+\varepsilon$. Therefore, the total noise referred to the channel input can be calculated by
\begin{equation}
\begin{aligned}
\chi_{tot}&=\chi_{line}+\chi_{hom}/\eta \\
&=\frac{1+v_{el}}{\tau\eta}-1+\varepsilon.
\end{aligned}
\end{equation}
After passing the untrusted quantum channel, the covariance matrix $\Gamma_{AB_{3}}^{(j)}$ has the form as follows
\begin{equation}       
\Gamma_{AB_{3}}^{(j)}\!\!=\!\!
\left(                 
\begin{array}{cc}   
 a\mathbb{I} & c\sigma_{z}\\  
    c\sigma_{z} & b\mathbb{I}\\  
  \end{array}
\right)=    
\left(                 
 \begin{array}{cc}   
   X'\mathbb{I} & \sqrt{\eta}Z'_{4}\sigma_{z}\\  
    \sqrt{\eta}Z'_{4}\sigma_{z} & \eta(Y'+\chi_{line})\mathbb{I}\\  
  \end{array}
\right).            
\end{equation}
As a result, the Shannon mutual information between Alice and Bob, $I(A:B)$, can be calculated by
\begin{equation}
\begin{aligned}
I(A:B)&=\frac{1}{2}\mathrm{log_2}\frac{V_A}{V_{A|B}},
\end{aligned}
\end{equation}
where $V_A=(a+1)/2$, $V_B=b$ and
\begin{equation}
\begin{aligned}
V_{A|B}&=V_A-\frac{\eta Z_4'^2}{2V_B} \\
&=a-\frac{c^2}{2b}.
\end{aligned}
\end{equation}

After Bob applies homodyne measurement,  Eve purifies the whole system so that the mutual information between Eve and Bob can be expressed as
\begin{equation}
\begin{aligned}
S(E:B)&=S(E)-S(E|B) \\
&=S(AB)-S(A|B) \\
&=G[(\kappa_{1}-1)/2]+G[(\kappa_{2}-1)/2] \\
&-G[(\kappa_{3}-1)/2]-G[(\kappa_{4}-1)/2],
\end{aligned}
\end{equation}
where the Von Neumann entropy $G(x)$ is given by \begin{equation}
\begin{aligned}
G(x)=(x+1)\mathrm{log}_{2}(x+1)-x\mathrm{log}_{2}x
\end{aligned},
\end{equation}
and the symplectic eigenvalues $\kappa_{1,2,3,4}$ can be calculated by
\begin{equation}
\begin{aligned}
\kappa_{1,2}^{2}=\frac{1}{2}(A\pm\sqrt{A^{2}-4B})
\end{aligned},
\end{equation}
and 
\begin{equation}
\begin{aligned}
\kappa_{3,4}^2=\frac{1}{2}(C\pm\sqrt{C^2-4D})
\end{aligned},
\end{equation}
with 
\begin{equation}
\begin{aligned}
A&=V^2+\eta^2(V+\chi_{line})^2-2\eta Z_4'^2, \\
B&=\eta(V^2+V\chi_{line}-Z_4'^2)^2, \\
C&=\frac{A\chi_{hom}+V\sqrt{B}+\eta (V+\chi_{line})}{\eta (V+\chi_{tot})}, \\
D&=\sqrt{B}\frac{V+\sqrt{B}\chi_{hom}}{\eta (V+\chi_{tot})}.
\end{aligned}
\end{equation}

\section{Secret key rate in the finite-size scenario}

In the traditional CVQKD protocol, the secret key rate calculated by taking finite-size effect into account is expressed as \cite{Leverrier:2010es}
\begin{equation}\label{fini1}
\begin{aligned}
K_{fini}=\frac{n}{N}[\beta I(A:B)-S_{\epsilon_{PE}}(E:B)-\Delta(n)],
\end{aligned}
\end{equation}
where $\beta$ and $I(A:B)$ are as same as the afore-mentioned definitions, $N$ denotes the total exchanged signals and $n$ denotes the number of signals that is used for sharing key between Alice and Bob. The remained signals $m=N-n$ is  used for parameter estimation. $\epsilon_{PE}$ is the failure probability of parameter estimation and the parameter $\Delta(n)$ is related to the security of the privacy amplification, which is given by
\begin{equation}
\begin{aligned}
\Delta(n)=(2\mathrm{dim}\mathcal{H}_{B}+3)\sqrt{\frac{\mathrm{log}_2(2/\bar{\epsilon})}{n}}+\frac{2}{n}\mathrm{log}_2(1/\epsilon_{PA}),
\end{aligned}
\end{equation}
where $\bar{\epsilon}$ is a smoothing parameter, $\epsilon_{PA}$ is the failure probability of privacy amplification, and $\mathcal{H}_{B}$ is the Hilbert space corresponding to the Bob's raw key. Since the raw key is usually encoded on binary bits, we have $\mathrm{dim}\mathcal{H}_{B}=2$. For the proposed long-distance CVQKD scheme, the secret key rate in Eq. (\ref{fini1}) can be rewritten as
\begin{equation}\label{fini}
\begin{aligned}
K_{fini}=\frac{nP_{(j)}^{\hat{\Pi}_1}}{N}[\beta \zeta_{opt}I(A:B)-S_{\epsilon_{PE}}(E:B)-\Delta(n)].
\end{aligned}
\end{equation}

In the finite-size scenario, $S_{\epsilon_{PE}}(E:B)$ needs to be calculated in parameter estimation procedure where one can find a covariance matrix $\Gamma_{\epsilon_{PE}}$ which minimizes the secret key rate with a probability of $1-\epsilon_{PE}$ and can be calculated by $m$ couples of correlated variables $(x_i,y_i)_{i=1\cdots m}$  in the following form
\begin{equation}       
\Gamma_{\epsilon_{PE}}\!\!=\!\! 
\left(                 
 \begin{array}{cc}   
   X'\mathbb{I} & tZ'_{4}\sigma_{z}\\  
    tZ'_{4}\sigma_{z} & (t^2 X'+\sigma^2)\mathbb{I}\\  
  \end{array}
\right),         
\end{equation}
where $t=\sqrt{\eta}$ and $\sigma^2=1+\eta(\varepsilon-3)$ are compatible with $m$ sampled data except with probability $\epsilon_{PE}/2$. The maximum-likelihood estimators $\hat{t}$ and $\hat{\sigma^2}$ respectively has the follow distributions
\begin{equation}
\begin{aligned}
\hat{t}\sim \big(t,\; \frac{\sigma^2}{\sum_{i=1}^mx_i^2}\big)\quad \mathrm{and} \quad \frac{m\hat{\sigma}^2}{\sigma^2}\sim\chi^2(m-1),
\end{aligned}
\end{equation}
where $t$ and $\sigma^2$ are the authentic values of the parameters. In order to maximize the value of the Holevo information between Eve and Bob with the statistics except with probability $\epsilon_{PE}$, we compute $t_{min}$ (the lower bound of $t$) and $\sigma_{max}^2$ (the upper bound of $\sigma^2$) in the limit of large $m$, namely
\begin{equation}
\begin{aligned}
t_{min}&=\sqrt{\eta}-z_{\epsilon_{PE}/2}\sqrt{\frac{1+\eta(\varepsilon-3)}{mX'}}, \\
\sigma_{max}^2&=1+\eta(\varepsilon-3)+z_{\epsilon_{PE}/2}\frac{\sqrt{2}[1+\eta(\varepsilon-3)]}{\sqrt{m}},
\end{aligned}
\end{equation}
where $z_{\epsilon_{PE}/2}$ is such that $1-\mathrm{erf}(z_{\epsilon_{PE}/2}/\sqrt{2})/2=\epsilon_{PE}/2$ and erf is the error function defined as
\begin{equation}
\begin{aligned}
\mathrm{erf}(x)=\frac{2}{\pi}\int_0^xe^{-t^2}\mathrm{d}t.
\end{aligned}
\end{equation}
The above-mentioned error probabilities can be set to
\begin{equation}
\begin{aligned}
\bar{\epsilon}=\epsilon_{PE}=\epsilon_{PA}=10^{10}.
\end{aligned}
\end{equation}
Finally, one can calculate the secret key rate in the finite-size scenario using the derived bounds $t_{min}$ and $\sigma_{max}^2$.

\begin{table}
\caption{The parameters of the proposed scheme in the composable security framework}
\label{tab:1}       
\begin{tabular}{ll}
\hline\noalign{\smallskip}
\it{parameter} &  \it{definition}  \\
\noalign{\smallskip}\hline\noalign{\smallskip}
\hline
$N$ & total number of exchanged light pulses. \\
 \hline
$n$ & size of final key if the protocol did not abort.  \\
\hline
$d$ & number of bits on which each measurement\\
 & result is encoded. \\
 \hline
$leak_{EC}$ & size of Bob's communication to Alice during \\
 & error correction step.\\
\hline
$\epsilon_{PE}$ & maximum failure probability of parameter\\
& estimation step. \\
\hline
$\epsilon_{cor}$ & small probability of the failure that the keys of \\
 & Alice and Bob do not identical and the protocol \\
 &did not abort.\\
 \hline
$n_{PE}$ & number of bits that Bob sends to Alice during \\
& parameter estimation step.\\
\hline
$\Omega_{a}^{max}$,$\Omega_{b}^{max}$, & bounds on covariance matrix elements, which \\
$\Omega_{c}^{min}$ & must be apt in the realization of the protocol.\\
\noalign{\smallskip}\hline
\end{tabular}
\end{table}

\section{Secret key rate of the CVQKD in composable security}

We detail the generation of secret key rate of the proposed long-distance CVQKD scheme provided by composable security framework. In Tab. \ref{tab:1}, we show the definition of parameters in the composable security case. Before the calculation, we give a theorem of composable security for the proposed  scheme \cite{Leverrier:2015he}.

The proposed long-distance CVQKD protocol is $\epsilon$-secure against collective attacks if $\epsilon=2\epsilon_{sm}+\overline{\epsilon}+\epsilon_{PE}/\epsilon+\epsilon_{cor}/\epsilon+\epsilon_{ent}/\epsilon$ and if the final key length $n$ is chosen such that
\begin{equation}
\begin{aligned}
n&\leq2N\hat{H}_{MLE}(U)-NF(\Omega_{a}^{max},\Omega_{b}^{max},\Omega_{c}^{min})\\
&-leak_{EC}-\Delta_{AEP}-\Delta_{ent}-2\log\frac{1}{2\overline{\epsilon}},
\end{aligned}
\end{equation}
where $\hat{H}_{MLE}(U)$ is the empiric entropy of $U$, the maximum likelihood estimator (MLE) of $H(U)$ to be $\hat{H}_{MLE}(U)=-\sum_{i=1}^{2^{d}}\hat{p}_{i}\log\hat{p}_{i}$ with $\hat{p}_{i}=\frac{\hat{n}_{i}}{dN}$ denotes the relative frequency of obtaining the value $i$, and $\hat{n}_{i}$ is the number of times the variable $U$ takes the value $i$ for $i\in\{1,\cdots,2^{d}\}$,  $F$ is the function computing the Holevo information between Eve and Bob, and
\begin{equation}
\begin{aligned}
\Delta_{AEP} &=\sqrt{N}(d+1)^2+\sqrt{16N}(d+1)\log_{2}\frac{2}{\epsilon_{sm}^{2}} \\
&+\sqrt{4N}\log_{2}\frac{2}{\epsilon^{2}\epsilon_{sm}}-4\frac{\epsilon_{sm}d}{\epsilon},
\end{aligned}
\end{equation}
\begin{equation}
\begin{aligned}
\Delta_{ent}=\log_{2}\frac{1}{\epsilon}-\sqrt{4N\log^{2}(2N)\log(2/\epsilon_{sm})}
\end{aligned}.
\end{equation}

Now, we consider the calculation of secret key rate of the proposed long-distance CVQKD scheme provided by composable security framework. Since the transmission channel is characterized by transmissivity $\eta$ and excess noise $\varepsilon$, the following model is used for error correction 
\begin{equation}
\begin{aligned}
\beta I(A:B)=2\hat{H}_{MLE(U)}-\frac{1}{2n}leak_{EC},
\end{aligned}
\end{equation}
where $I(A:B)$ represents the mutual information between Alice and Bob, $\beta$ denotes the reconciliation efficiency. For the proposed protocol, we obtain 
\begin{equation}
\begin{aligned}
I(A:B)&=\frac{1}{2}\log_{2}(1+SNR)\\
&=\frac{1}{2}\log_{2}\left(1+\frac{\eta V_{M}}{2+\eta\varepsilon}\right).
\end{aligned}
\end{equation}
Moreover, assuming that the success probability of parameter estimation is at least $0.99$, and hence the robustness of the proposed protocol is $\epsilon_{rob}\leq10^{-2}$. Consequently, the values of random variables $\left|\left| X\right|\right|^{2}$, $\left|\left| Y\right|\right|^{2}$ and $\langle X,Y\rangle$ satisfy the following restraints 
\begin{equation}
\begin{aligned}
\left|\left| X\right|\right|^{2}\leq(N+3\sqrt{N})X', \\
\end{aligned}
\end{equation}
\begin{equation}
\begin{aligned}
\left|\left| Y\right|\right|^{2}&\leq\eta(N+3\sqrt{N})(Y'+\chi_{line}),\\
\end{aligned}
\end{equation}
\begin{equation}
\begin{aligned}
\langle X,Y\rangle\geq(N-3\sqrt{N})\sqrt{\eta}Z_4',
\end{aligned}
\end{equation}
The above-mentioned restraints can be achieved from the covariance matrix $\Gamma_{AB_{3}}^{(j)}$ of the proposed CVQKD scheme. According to these bounds, we have the definations 
\begin{equation}\label{oa}
\Omega_{a}^{max}=\frac{\left|\left| X\right|\right|^{2}}{N}\left[1+2\sqrt{\frac{\log(36/\epsilon_{PE})}{N/2}}\right]-1,
\end{equation}
\begin{equation}\label{ob}
\Omega_{b}^{max}=\frac{\left|\left| Y\right|\right|^{2}}{N}\left[1+2\sqrt{\frac{\log(36/\epsilon_{PE})}{N/2}}\right]-1,
\end{equation}
\begin{equation}\label{oc}
\Omega_{c}^{min}=\frac{\langle X,Y\rangle}{N}-5(\left|\left| X\right|\right|^{2}+\left|\left| Y\right|\right|^{2})\sqrt{\frac{\log(8/\epsilon_{PE})}{{(N/2)}^{3}}}.
\end{equation}
Finally, we can calculate the secret key rate of the proposed scheme provided by composable security as follows
\begin{widetext}
\begin{equation}
\begin{aligned}
K_{comp}=P_{(j)}^{\hat{\Pi}_1}(1-\epsilon_{rob})\{\beta \zeta_{opt}I(A:B)
-F(\Omega_{a}^{max}, \Omega_{b}^{max}, \Omega_{c}^{min})
-\frac{1}{N}(\Delta_{AEP}+\Delta_{ent}+2\log_{2}\frac{1}{2\overline{\varepsilon}})\}.
\end{aligned}
\end{equation}
\end{widetext}

In addition, we should optimize over all parameters compatible with $\epsilon=10^{-20}$. However, in order to simplify the data process, we make the following choices \begin{equation}
\begin{aligned}
\epsilon_{sm}&=\overline{\epsilon}=10^{-21}, 
\epsilon_{PE}&=\epsilon_{cor}=\epsilon_{ent}=10^{-41}.
\end{aligned}
\end{equation} which slightly sub-optimizes the performance of the proposed CVQKD protocol \cite{Leverrier:2015he}.

\end{appendix}


\bibliography{mybibfile}

\end{document}